\RequirePackage{ifpdf}
\ifpdf 
\documentclass[pdftex]{sigma}
\else
\documentclass{sigma}
\fi

\begin{document}

\renewcommand{\PaperNumber}{011}

\FirstPageHeading

\ShortArticleName{Connections Between Symmetries and Conservation Laws}

\ArticleName{Connections Between Symmetries\\ and Conservation Laws}

\Author{George BLUMAN}
\AuthorNameForHeading{G. Bluman}

\Address{Department of Mathematics, University of British Columbia, 
Vancouver, BC, Canada V6T1Z2} 
\Email{\href{mailto:bluman@math.ubc.ca}{bluman@math.ubc.ca}} 

\ArticleDates{Received July 29, 2005, in final form October 22, 2005; Published online October 26, 2005}

\Abstract{This paper presents recent work on connections between symmetries and conservation laws. 
After reviewing Noether's
theorem and its limitations, we present the Direct Construction Method to show 
how to find directly the conservation laws
for any given system of differential equations. This method yields the multipliers for conservation laws as well as an
integral formula for corresponding conserved densities. The action of a symmetry 
(discrete or continuous) on a conservation
law yields conservation laws. Conservation laws yield non-locally related systems that, in turn, can yield nonlocal
symmetries and in addition be useful for the application of other mathematical methods. From its admitted symmetries or
multipliers for conservation laws, one can determine whether or not a given system of differential equations can be
linearized by an invertible transformation.}


\Classification{58J70; 58J72; 70G65; 70G75; 70H03; 70H33; 70S10}

\section{Introduction}\label{sec1}

In this paper, we discuss several connections between symmetries and conservation laws. In the classical Noether's theorem
\cite{Noether}, if a given system of differential equations (DEs) has a variational principle, then a continuous symmetry
(point, contact or higher order) that leaves invariant the action functional to within a divergence yields a conservation
law \cite{Bessel_Hagen, Ibragimov, Olver, Bluman&Kumei}. A system of DEs (as written) has a variational principle if and only if its
linearized system is self-adjoint \cite{Olver}. There are several limitations to Noether's theorem: It is restricted to
variational systems. Consequently, to be applicable to a given system as written, the given system must be of even order,
have the same number of dependent variables as the number of equations in the system and have no dissipation. There is also
the difficulty of finding symmetries admitted by the action functional. Moreover, the use of Noether's theorem to find
conservation laws is coordinate-dependent.

The Direct Construction Method (DCM) gets around the limitations of Noether's theorem for finding 
conservation laws. A given
system of DEs has a conservation law if and only if all Euler operators associated with its 
dependent variables annihilate
the scalar product of multipliers with each DE where in the resulting expression the solutions 
of the system of DEs are
replaced by arbitrary functions. The DCM \cite{Anco&Bluman,Anco&Bluman2,Anco&Bluman3,Anco&Bluman4} 
yields an integral formula for the
corresponding conservation law for any such set of multipliers. On the solution space of the given system of DEs,
multipliers are symmetries provided its linearized system is self-adjoint; otherwise they are solutions of the adjoint of
the linearized system. The use of adjoint linearization to find multipliers for conservation laws is extensively discussed
in \cite{Kras} and references therein.  A~comprehensive discussion of the computational aspects of the DCM and comparisons
with three other approaches to calculate conservation laws appears in \cite{Wolf}.

For any system of DEs, a symmetry (point or contact) of the system maps conservation laws to conservation laws. In
particular, any symmetry of the system induces a symmetry that leaves invariant the determining equations for multipliers of
its conservation laws. Moreover, there is a formula to determine the action of a symmetry (discrete or continuous) on the
multipliers of a given conservation law and from this action one can determine in advance whether or not the resulting
conservation law will be new \cite{Bluman&Temuerchaolu&Anco}.

For any system of DEs, useful conservation laws yield equivalent non-locally related systems of DEs \cite{Bluman&DoranWu}.
Such non-locally related systems (and some of their subsystems) in turn can yield nonlocal conservation laws and nonlocal
symmetries for the given system. This leads to the important problem of finding extended trees of equivalent non-locally
related systems of DEs~\cite{Bluman&Cheviakov}.

For any system of DEs, from its admitted symmetry group one can determine whether or not it can be mapped into a linear
system and find the explicit mapping \cite{Bluman&Kumei,Kumei&Bluman2,Bluman&Kumei3,Bluman&Kumei4}; from its admitted
multipliers for conservation laws, one can also determine whether or not it can be mapped into a linear system--here the
solution of the determining system for multipliers yields the adjoint of the resulting linear system
\cite{Bluman&DoranWu,Anco&Wolf&Bluman}.

This paper is organized as follows. In section two we review Noether's theorem and discuss its limitations. In section three
we present the Direct Construction Method to find conservation laws. In section four we consider the use of symmetries to
find new conservation laws from known conservation laws. In section five we show how conservation laws lead to equivalent
non-locally related systems for finding nonlocal symmetries. In section six we compare the roles of symmetries and
multipliers for conservation laws in determining whether or not a given system can be linearized by an invertible
transformation. Finally, several examples are presented in section seven to illustrate the work discussed in this paper.

\section{Review of Noether's theorem}\label{sec2}

Consider a functional $J[U]$ in terms of $n$ independent variables $x=(x_1 ,\ldots ,x_n )$ and $m$ arbitrary functions
$U=(U^1(x),\ldots ,U^m(x))$ of the independent variables $x$ defined on a domain $\Omega$,
\begin{gather}
\label{eq1} J[U]=\int_\Omega {L[U]dx=\int_\Omega {L(x,U,\partial U,\ldots ,\partial ^kU)dx} } ,
\end{gather}
with $x$ derivatives of $U$ up to some order $k$. We let $\partial U,\;\partial ^2U,$ etc. 
denote all derivatives of~$U^\gamma $ of a given order with respect to $x_i .$ We 
denote partial derivatives $\partial /\partial x_i $ by subscripts~$i$, 
i.e., $U_i^\gamma ={{\partial U^\gamma } \over {\partial x_i }},\;U_{ij}^\gamma ={{\partial
^2U^\gamma } \over {\partial x_i \partial x_j }},$ etc. Corresponding total 
derivatives are denoted by $D_i
={\partial \over {\partial x_i }}+U_i^\gamma {\partial \over {\partial U^\gamma }}+\cdots +U_{ii_1
\cdots i_j }^\gamma {\partial \over {\partial U_{i_1 \cdots i_j }^\gamma }}+\cdots $. 
We let $F[U]$ denote a
function $F $ depending on $x$, $U$ and derivatives of $U$ with respect to $x$. 
Throughout this paper we use the summation
convention for any repeated indices. The \textit{Euler operator} with respect to $U^\gamma$ is denoted by
\[
E_{U^\gamma } =\frac{\partial }{\partial U^\gamma }-D_i \frac{\partial }{\partial U_i^\gamma }+\cdots +(-1)^jD_{i_1 } \cdots
D_{i_j } \frac{\partial }{\partial U_{i_1 \cdots i_j }^\gamma }+\cdots .
\]

From the calculus of variations, it follows that if $U=u$ is an extremum of $J[U],$ then
\begin{gather}
\label{eq2} \left. {\left( {E_{U^\gamma } L[U]} \right)} \right|_{U=u} =0,\qquad \gamma =1,\ldots ,m.
\end{gather}
The function $L[U]$ is called a \textit{Lagrangian} and the integral $J[U]$ of (\ref{eq1}) an \textit{action} (or
\textit{variational})\textit{ integral}. The extremal system of differential equations (\ref{eq2}) with independent
variables $x$ and dependent variables $u=(u^1(x),\ldots ,u^m(x))$ are called the \textit{Euler--Lagrange equations} 
for an extremum $U=u(x)$ of $J[U].$

Now consider the Euler--Lagrange equations (\ref{eq2}) as a given system of differential equations
\begin{gather}
\label{eq3} G_\gamma [u]=E_{u^\gamma } L[u]=0,\qquad \gamma =1,\ldots ,m
\end{gather}
whose solutions $u(x)$ are extrema of some action integral (\ref{eq1}). A \textit{local conservation law}
of system~(\ref{eq3}) is a divergence expression
\begin{gather}
\label{eq4} D_i \Phi ^i[u]=0
\end{gather}
that holds for all solutions $u(x)$ of system (\ref{eq3}); $\Phi ^i[u],\;i=1,\ldots ,n$ are called the \textit{conserved
densities.}

Consider a local symmetry generator in evolutionary form \cite{Ibragimov,Olver,Bluman&Kumei}
\begin{gather}
\label{eq5} X=\eta ^\sigma [U]\frac{\partial }{\partial U^\sigma };
\end{gather}
$X$ defines a \textit{point symmetry }generator if $\eta ^\sigma [U]$ is of the linear 
form $\eta ^\sigma [U]=\eta ^\sigma
(x,U)+\xi _i (x,U)U_i^\sigma$, $\sigma =1,\ldots ,m$; 
$X$ defines a \textit{contact symmetry} generator if $m=1$ and $\eta
^1[U]=\eta [U]$ depends at most on first partials of $U$, i.e.~$\eta [U]=\eta (x,U,\partial U)$; $X$ defines a
\textit{higher order symmetry }generator if it is neither a point symmetry nor contact symmetry generator. Higher order
symmetries were first mentioned by Emmy Noether in her celebrated paper \cite{Noether}. The $p\mbox{th}$ extension of a
local symmetry generator (\ref{eq5}) is given by
\[
X^{(p)}=\eta ^\sigma [U]{\partial \over {\partial U^\sigma }}+(D_i \eta ^\sigma [U]){\partial \over
{\partial U_{^i}^\sigma }}+\cdots + (D_{i_1 } \cdots D_{i_p } \eta ^\sigma [U]){\partial \over {\partial U_{^{i_1
\cdots i_p }}^\sigma }}.
\]
A local symmetry generator (\ref{eq5}) defines a \textit{variational symmetry }of the action integral (\ref{eq1}) if and
only if
\begin{gather}
\label{eq6} X^{(k)}L[U]=D_i f^i[U]
\end{gather}
holds for some functions $f^i[U].$ If (\ref{eq5}) defines a variational symmetry of (\ref{eq1}), then one can derive the
identity \cite{Ibragimov,Olver,Bluman&Kumei}
\begin{gather}
\label{eq7} \eta ^\sigma [U]G_\sigma [U]=\eta ^\sigma [U]E_{U^\sigma } L[U]=D_i (f^i[U]-W^i[U,\eta [U]])
\end{gather}
where
\[
W^i[U,\eta [U]]=\left( {\eta ^\sigma [U]E_{U_i^\sigma } +\sum\limits_{j=2} {(D_{i_1 } \cdots D_{i_{j-1} } \eta ^\sigma
[U])E_{U_{ii_1 \cdots i_{j-1} }^\sigma } } } \right)L[U].
\]
Consequently, a variational symmetry (\ref{eq5}) of the action integral (\ref{eq1}) yields the conservation law
\[
D_i \Phi ^i[u]=D_i (W^i[u,\eta [u]]-f^i[u])=0
\]
of the given system of differential equations arising from the extrema of the action integral (\ref{eq1}). In particular,
from (\ref{eq7}) we see that Noether's theorem yields a conservation law of the variational system (\ref{eq3}) for any set
of \textit{multipliers }$\{\eta ^\sigma [U]\}$ for which $\eta ^\sigma [U]{\partial \over {\partial U^\sigma }}$
is a variational symmetry of the action integral (\ref{eq1}).

Since a variational symmetry leaves the Lagrangian $L[U]$ invariant to within a divergence, it follows that a variational
symmetry of the action integral $J[U]$ leaves invariant its family of extrema $U=u.$ Hence a variational symmetry
(\ref{eq5}) of the action integral (\ref{eq1}) yields a symmetry $X=\eta ^\sigma [u]{\partial \over {\partial
u^\sigma }}$ of the system of differential equations (\ref{eq3}), i.e.,
\begin{gather}
\label{eq8} \left. {X^{(2k)}G_\gamma [U]} \right|_{U=u} =0,\qquad \gamma =1,\ldots ,m.
\end{gather}
Conversely, a symmetry $X=\eta ^\sigma [u]{\partial \over {\partial u^\sigma }}$ of a system of DEs (\ref{eq3})
derivable from an action integ\-ral~(\ref{eq1}) does not necessarily yield a variational symmetry $X=\eta ^\sigma
[U]{\partial \over {\partial U^\sigma }}$ of the action integ\-ral~(\ref{eq1}).

\subsection{Limitations of Noether's theorem}\label{s2_1}

\begin{enumerate}
\itemsep=0pt
\item \textit{The difficulty of finding variational symmetries}. 
To find variational symmetries for a given variational system (\ref{eq3}), 
first one determines local symmetries $X=\eta ^\sigma [u]{\partial \over {\partial u^\sigma }}$ 
admitted by the Euler--Lagrange equations (\ref{eq3}) through the extension 
of Lie's algorithm to include higher order symmetries. 
Then for each such admitted symmetry, one checks if 
$X=\eta ^\sigma [U] {\partial \over {\partial U^\sigma }}$ 
leaves invariant the Lagrangian $L[U]$ to within a divergence, i.e., equation (\ref{eq6}) is satisfied.
\item \textit{A given system of DEs is not variational as written}. 
A given system of differential equations, as written, is variational if and only 
if its linearized system (Fr\'{e}chet derivative) is self-adjoint \cite{Olver}. 
Consequently, it is necessary that a given system of DEs, as written, 
must be of even order, have the same number of equations in the system 
as its number of dependent variables and be non-dissipative to directly admit a variational principle.
\item \textit{Artifices may make a given system of DEs variational}. Such artifices include:
    \begin{itemize}
\itemsep=0pt
    \item The use of multipliers. As an example, the PDE
\[
u_{tt} +{H}'(u_x )u_{xx} +H(u_x )=0,
\]
as written, does not admit a variational principle since its linearized equation $\varsigma _{tt} +{H}'(u_x )\varsigma _{xx}
+({H}''(u_x )+{H}'(u_x ))\varsigma _x =0$ is not self-adjoint. However, the equivalent PDE
\[
e^x[u_{tt} +{H}'(u_x )u_{xx} +H(u_x )]=0,
\]
as written, is self-adjoint!

\item The use of a contact transformation of the variables. As an example, the ODE
\begin{gather}
\label{eq9} {y}''+2{y}'+y=0,
\end{gather}
as written, obviously does not admit a variational principle. But the point transformation $x\to X=x$,
$y\to Y=ye^x,$ maps ODE
(\ref{eq9}) into the variational ODE ${Y}''=0.$ It is well-known 
that every second order ODE, written in solved form, can be
mapped into ${Y}''=0$ by some contact transformation but 
there is no finite algorithm to find such a transformation.

\item The use of a differential substitution. As an example, the Korteweg--de Vries (KdV) equation
\[
u_{xxx} +uu_x +u_t =0,
\]
as written, obviously does not admit a variational principle since 
it is of odd order. But the well-known differential
substitution $u=v_x $ yields the related transformed KdV equation 
$v_{xxxx} +v_x v_{xx} +v_{xt} =0$ which is the
Euler--Lagrange equation for an extremum $V=v$ of the action integral 
with Lagrangian $L[V]={1 \over 2}(V_{xx}
)^2-{1 \over 6}(V_x )^3-{1 \over 2}V_x V_t .$

    \end{itemize}

\item \textit{Noether's theorem is coordinate-dependent.} The use of Noether's
 theorem to obtain a~con\-ser\-vation law is coordinate-dependent since the action 
 of a contact transformation can transform a DE having 
 a variational principle to one that does not have one. 
 On the other hand it is well-known that conservation laws are 
 coordinate-independent in the sense that a contact transformation maps 
 a conservation law into a conservation law (see, for examp\-le~\cite{Bluman&Temuerchaolu&Anco}.)
\item \textit{Artifice of a Lagrangian}.
 One should be able to find the conservation laws of 
 a given system of DEs directly without the need to find
  a related functional whether or not the given system is variational.

\end{enumerate}

\section{Direct construction method to find conservation laws}\label{sec3}

Through the Direct Construction Method, one can get around the limitations of Noether's theorem. Here one works directly
with a given system of DEs whether or not its linearized system is self-adjoint. The starting point is 
the observation from
equation (\ref{eq7}) that conservation laws obtained through Noether's theorem arise from multipliers that must be local
symmetries of the given system of DEs (\ref{eq3}). Moreover (from equations (\ref{eq8})), 
the local symmetries of a system of DEs
are the solutions of its linearized system. Now suppose we have a given system of~$M$ differential equations
\begin{gather}
\label{eq10} G_\alpha [u]=G_\alpha \left(x,u,\partial u,\ldots ,\partial ^Ku\right)=0,\qquad \alpha =1,\ldots ,M
\end{gather}
with $n$ independent variables $x=(x_1 ,\ldots ,x_n )$ and 
$m$ dependent variables $u=(u^1,\ldots ,u^m).$ It is understood
that the system of DEs (\ref{eq10}) includes all its differential consequences. The local symmetry $X=\eta ^\sigma
[u]{\partial \over {\partial u^\sigma }}$ is admitted by (\ref{eq10}) if and only if
\begin{gather}
\label{eq11} \left. {X^{(K)}G_\alpha [u]} \right|_{G_\gamma [u]=0,\;\gamma =1,\ldots ,M} =0,\qquad \alpha =1,\ldots ,M.
\end{gather}
Equations (\ref{eq11}) are commonly called the linear \textit{determining equations }for the local symmetries admitted by
the given system (\ref{eq10}).\ Let L be the linearizing operator (Fr\'{e}chet derivative) for system (\ref{eq10}) that
arises from equations (\ref{eq11}). Then the determining equations (\ref{eq11}) can be expressed as a linear system
\begin{gather}
\label{eq12} \mbox{L}\eta [u]=0.
\end{gather}
The given system (\ref{eq10}), as written, has a variational principle if and only if L is a self-adjoint operator, i.e.,
$\mbox{L}^\ast =\mbox{L}$ where $\mbox{L}^\ast $ is the \textit{adjoint} of L.

A set of \textit{multipliers} $\{\Lambda ^\sigma [U]\}$ yields a conservation law of the given system (\ref{eq10}) if and
only if
\[
\Lambda ^\sigma [U]G_\sigma [U]=D_i \Phi ^i[U]
\]
for some $\{\Phi ^i[U]\}.$ Then the conservation law $D_i \Phi ^i[u]=0$ holds if $U=u$ solves the given system~(\ref{eq10}).
The aim is to first find all such sets of multipliers $\{\Lambda ^\sigma [U]\}$ for conservation laws of~(\ref{eq10}) and
then obtain the corresponding set of conserved densities $\{\Phi ^i[u]\}$ for each set of multipliers. The determining
equations for multipliers result from the following theorem that can be proved by direct computation.

\begin{theorem}\label{theorem1}
A set of multipliers $\{\Lambda ^\sigma [U]\}$ yields a conservation law of the given system of differential equations
\eqref{eq10} if and only if the equations
\begin{gather}
\label{eq13} E_{U^\gamma } (\Lambda ^\sigma [U]G_\sigma [U])=0,\quad \gamma =1,\ldots ,m
\end{gather}
{hold for} arbitrary {functions} $U=(U^1(x),\ldots ,U^m(x)).$

\end{theorem}

The set of equations (\ref{eq13}) are the linear determining equations for the multipliers for conservation laws 
admitted by
the given system (\ref{eq10}). 
One can show \cite{Olver,Anco&Bluman,Anco&Bluman2,Anco&Bluman3,Anco&Bluman4,Anco&Bluman5} that
if $\{\Lambda ^\sigma [U]\}$ yields a conservation law for system (\ref{eq10}), 
then for the adjoint operator $\mbox{L}^\ast
$ of the linearization operator L arising from the linear determining 
equations (\ref{eq12}) for local symmetries of system~(\ref{eq10}), one has
\[
\mbox{L}^\ast \Lambda [u]=0.
\]
If $\mbox{L}^\ast =\mbox{L}$, then $\Lambda ^\sigma [u]{\partial \over {\partial u^\sigma }}$ is a local symmetry
admitted by system (\ref{eq10}); if $\mbox{L}^\ast \ne \mbox{L,}$ then $\Lambda ^\sigma [u]{\partial \over
{\partial u^\sigma }}$ is not a local symmetry admitted by system (\ref{eq10}). When $\mbox{L}^\ast =\mbox{L}$, the
determining equations (\ref{eq13}) \textit{include as a subset} the set of determining equations for local symmetries
admitted by system~(\ref{eq10}) and in this case any solution of the determining equations (\ref{eq13}) must be a
variational symmetry. Most importantly, given a set of multipliers $\{\Lambda ^\sigma [U]\}$ 
yielding a conservation law for
system (\ref{eq10}), the integral formulas for the corresponding conserved densities $\{\Phi ^i[u]\}$ are given in
\cite{Anco&Bluman,Anco&Bluman2,Anco&Bluman3,Anco&Bluman4,Anco&Bluman5}.

\section{Use of symmetries to find new conservation laws from known conservation laws}\label{sec4}

Any symmetry (discrete or continuous) admitted by a given system of DEs (\ref{eq10}) maps a conservation law of (\ref{eq10})
into another conservation law of (\ref{eq10}). Usually, the same conservation law of (\ref{eq10}) is obtained. An admitted
symmetry of system (\ref{eq10}) induces a symmetry that leaves invariant the linear determining system (\ref{eq13}) for its
multipliers. Hence, it follows that if we determine the action of a symmetry on a set of multipliers $\{\Lambda ^\sigma
[U]\}$ for a known conservation law of (\ref{eq10}) to obtain another set of multipliers $\{\hat {\Lambda }^\sigma [U]\}$,
then a priori we see whether or not a~new conservation is obtained for (\ref{eq10}).

Suppose the invertible point transformation
\begin{gather}
\label{eq14} x=x(\tilde {x},\tilde {U}),\qquad U=U(\tilde {x},\tilde {U}),
\end{gather}
with inverse
\[
\tilde {x}=\tilde {x}(x,U),\qquad \tilde {U}=\tilde {U}(x,U),
\]
is admitted by a given system of DEs (\ref{eq10}). Then
\begin{gather}
\label{eq15} G_\alpha [U]=A_\alpha ^\beta [\tilde {U}]G_\beta [\tilde {U}]
\end{gather}
holds for some $\{A_\alpha ^\beta [\tilde {U}]\}.$ One can easily prove the following theorem
\cite{Bluman&Temuerchaolu&Anco,Popo}.

\begin{theorem}\label{theorem2}
{Under the point transformation} \eqref{eq14}, {there exist functions} $\Psi ^i[\tilde {U}]${ such that}
\begin{gather}
\label{eq16} J[\tilde {U}]D_i \Phi ^i[U]=\tilde {D}_i \Psi ^i[\tilde {U}]
\end{gather}
{where the Jacobian determinant}
\begin{gather}
\label{eq17} J[\tilde {U}]=\frac{D(x_1 ,\ldots ,x_n )}{D(\tilde {x}_1 ,\ldots ,\tilde {x}_n )}=\left|
{{\begin{array}{*{20}c}
 {\tilde {D}_1 x_1 } \hfill & {\tilde {D}_1 x_2 } \hfill & \cdots \hfill &
{\tilde {D}_1 x_n } \hfill \\
 {\tilde {D}_2 x_1 } \hfill & {\tilde {D}_2 x_2 } \hfill & \cdots \hfill &
{\tilde {D}_2 x_n } \hfill \\
 \vdots \hfill & \vdots \hfill & \vdots \hfill & \vdots \hfill \\
 {\tilde {D}_n x_1 } \hfill & {\tilde {D}_n x_2 } \hfill & \cdots \hfill &
{\tilde {D}_n x_n } \hfill \\
\end{array} }} \right|
\end{gather}
{and}
\begin{gather}
\label{eq18} \Psi ^{i_1 }[\tilde {U}]=\pm \left| {{\begin{array}{*{20}c}
 {\Phi ^1[U]} \hfill & {\Phi ^2[U]} \hfill & \cdots \hfill & {\Phi ^n[U]}
\hfill \\
 {\tilde {D}_{i_2 } x_1 } \hfill & {\tilde {D}_{i_2 } x_2 } \hfill & \cdots
\hfill & {\tilde {D}_{i_2 } x_n } \hfill \\
 \vdots \hfill & \vdots \hfill & \vdots \hfill & \vdots \hfill \\
 {\tilde {D}_{i_n } x_1 } \hfill & {\tilde {D}_{i_n } x_2 } \hfill & \cdots
\hfill & {\tilde {D}_{i_n } x_n } \hfill \\
\end{array} }} \right|.
\end{gather}
{In} \eqref{eq18}, $(i_1 ,i_2 ,\ldots ,i_n )$ {denotes all cyclic permutations of the indices }$1,2,\ldots ,n${ and the
}$\pm $ {sign is chosen according to whether the permutation has even or odd order, i.e. }$(i_1 ,i_2 ,\ldots ,i_n
)=(1,2,\ldots ,n),\,(2,3,\ldots ,1),\ldots ,(n,1,\ldots ,n-1)${ are alternately even and odd. The coordinates of }$\Phi
^i[U]${ in} \eqref{eq16} {are expressed in terms of the point transformation} \eqref{eq14}.
\end{theorem}

The following corollary results immediately from Theorem 2.

\begin{corollary}\label{corollary1}
{If the point transformation} \eqref{eq14} {is admitted by the DE system} \eqref{eq10}, {then a conservation law}
\eqref{eq4} {of system} \eqref{eq10} {is mapped to a conservation law}
\[
D_i \Psi ^i[u]=0
\]
\textit{of system} \eqref{eq10} \textit{with conserved densities} $\Psi ^i[u].$
\end{corollary}

In \cite{Ibragimov, Olver, Stephani} it is shown how to obtain a conservation law from a known conservation law through the
action of the infinitesimal generator of an admitted continuous symmetry (in evolutionary (characteristic) form) on the
given conservation law.  Here it is necessary to use dif\-fe\-ren\-tial consequences of the given system (\ref{eq10}) even in the
case when the symmetry is equivalent to a point or contact symmetry.  A priori, one is unable to determine whether or not a
new conservation law is obtained.

 Now suppose the set of multipliers $\{\Lambda ^\sigma [U]\}$ yields a conservation law with densities $\Phi ^i[u]$ of a
given system of DEs (\ref{eq10}), i.e.,
\begin{gather}
\label{eq19} \Lambda ^\sigma [U]G_\sigma [U]=D_i \Phi ^i[U].
\end{gather}
One can prove the following theorem \cite{Bluman&Temuerchaolu&Anco}.

\begin{theorem}\label{theorem3}
{Suppose the point transformation} \eqref{eq14} {is admitted by} \eqref{eq10}
 {and }$\{\Lambda ^\sigma [U]\}$ {is a set of
multipliers for a conservation law of} \eqref{eq10} {with conserved densities }$\Phi ^i[u].${ Then}
\begin{gather}
\label{eq20} \hat {\Lambda }^\beta [\tilde {U}]G_\beta [\tilde {U}]=\tilde {D}_i \Psi ^i[\tilde {U}]
\end{gather}
{where}
\begin{gather}
\label{eq21} \hat {\Lambda }^\beta [\tilde {U}]=J[\tilde {U}]A_\alpha ^\beta [\tilde {U}]\Lambda ^\alpha [U],\qquad \beta
=1,\ldots ,M,
\end{gather}
{with the coordinates of derivative terms in }$\Lambda ^\alpha [U]${ expressed in terms of the natural extension of
transformation} \eqref{eq14} {to the derivatives of U} {\rm \cite{Ibragimov,Olver,Bluman&Kumei,Anco&Bluman5}}. 
In \eqref{eq20}, $\Psi ^i[\tilde {U}]$ {are given by the determinant} \eqref{eq18} {and, in} \eqref{eq21}, $A_\alpha ^\beta
[\tilde {U}]${ are given by} \eqref{eq15} {and the Jacobian determinant} $J[\tilde {U}]${ is given by} \eqref{eq17}.
\end{theorem}

After replacing the coordinates $\tilde {x}_i$ by $x_i ,\;\tilde {U}^\sigma$ by $U^\sigma ,\;\tilde {U}_i^\sigma$ by
$U_i^\sigma,$ etc. in (\ref{eq20}), one obtains the following corollary.

\begin{corollary}\label{corollary2}
{If }$\{\Lambda ^\alpha [U]\}${ is a set of multipliers yielding a conservation law of system}
\eqref{eq10} {admitting the
point transformation} \eqref{eq14}, {then} $\{\hat {\Lambda }^\beta [U]\}$ 
{yields a set of multipliers for a conservation
law of} \eqref{eq10} {where }$\hat {\Lambda }^\beta [U]${ is given by} \eqref{eq21} {after replacing }$\tilde {x}_i $
{by} $x_i$, $\tilde {U}^\sigma$ by $U^\sigma$, $\tilde {U}_i^\sigma $ {by} $U_i^\sigma$,
etc. The set of multipliers $\{\hat {\Lambda }^\beta [U]\}$ yields a new 
conservation law of system \eqref{eq10} {if and only if this set is
}nontrivial {on all solutions} $U=u(x)$ {of system} \eqref{eq10}, i.e.  $\hat {\Lambda }^\beta [u] \neq c\Lambda ^\beta
[u]$, $\beta =1,\ldots ,M$, for some constant $c$.
\end{corollary}

Now suppose the point transformation (\ref{eq14}) is a one-parameter $(\varepsilon )$ Lie group of point transformations
\begin{gather}
\label{eq22} x=x(\tilde {x},\tilde {U};\varepsilon )=e^{\varepsilon \tilde {X}}\tilde {x},\qquad U=U(\tilde {x},\tilde
{U};\varepsilon )=e^{\varepsilon \tilde {X}}\tilde {U}
\end{gather}
in terms of its infinitesimal generator (and extensions) $\tilde {X}=\xi _j (\tilde {x},\tilde {U}){\partial \over
{\partial \tilde {x}_j }}+\eta ^\sigma (\tilde {x},\tilde {U}){\partial \over {\partial \tilde {U}^\sigma }}$. If
(\ref{eq19}) holds, then from (\ref{eq16}) and the Lie group properties of (\ref{eq22}), it follows that
\begin{gather}
\label{eq23} J[U;\varepsilon ]e^{\varepsilon X}(\Lambda ^\sigma [U]G_\sigma [U])=D_i \Psi ^i[U;\varepsilon ]
\end{gather}
in terms of the (extended) infinitesimal generator $X=\xi _j (x,U){\partial \over {\partial x_j }}+\eta ^\sigma
(x,U){\partial \over {\partial U^\sigma }}$.

Now assume that the Lie group of point transformations (\ref{eq22}) is admitted by a given system of DEs (\ref{eq10}). Then
for some $\{a_{\rho \alpha }^\beta [U]\},$ one has
\[
e^{\varepsilon X}G_\alpha [U]=\sum {\varepsilon ^\rho a_{\rho \alpha }^\beta G_\beta [U].}
\]
Then, after expanding both sides of equation
 (\ref{eq23}) in terms of power series in $\varepsilon ,$ one obtains an expression
of the form
\begin{gather}
\label{eq24} \sum {\varepsilon ^p\hat {\Lambda }^\sigma [U;p]G_\sigma [U]=
\sum {\varepsilon ^p} } \left. {D_i \left({1
\over {p!}}{{d^p} \over {d\varepsilon ^p}}\Psi ^i[U;\varepsilon ]\right)} \right|_{\varepsilon =0} .
\end{gather}
Corresponding to the sequence of sets of multipliers $\{\hat {\Lambda }^\sigma [U;p]\},\;p=1,2,\ldots ,$ arising in
expression~(\ref{eq24}), one obtains a sequence of conservation laws
\[
\left. {D_i \left({{d^p} \over {d\varepsilon ^p}}\Psi ^i[u;\varepsilon ]\right)} \right|_{\varepsilon =0} =0,\qquad
p=1,2,\ldots
\]
for system (\ref{eq10}) from its known conservation law $D_i \Phi ^i[u]=0.$

The results presented in this section extend to contact transformations in the case of a scalar DE
\cite{Bluman&Temuerchaolu&Anco}.

\section{Connections between conservation laws\\ and nonlocal symmetries}\label{sec5}

Through useful conservation laws, one can obtain a tree of non-locally related systems and subsystems for a given system of
DEs. Such non-locally related systems lead to the possibility of using Lie's algorithm to seek nonlocal symmetries for the
given system. For details of the work discussed in this section see \cite{Bluman&Kumei,Bluman&DoranWu,Bluman&Cheviakov} and
references therein.

In this section, let $x=(x_1 ,x_2 )=(t,x);\;u=(u^1,\ldots ,u^m).$ Consider a given system of PDEs
\begin{gather}
\label{eq25} G_\alpha [u]=G_\alpha (t,x,u,\partial u,\ldots ,\partial ^ku)=0,\qquad \alpha =1,\ldots ,M.
\end{gather}
Suppose $G_1 [u]$ is written as a conservation law:
\begin{gather}
\label{eq26} G_1 [u]=D_t T[u]+D_x X[u]=0.
\end{gather}
Then through the conservation law (\ref{eq26}), one can introduce an auxiliary potential dependent variable $v$. It is easy
to see that system (\ref{eq25}) is \textit{non-locally }equivalent to the potential system
\begin{gather}
 v_t +X[u]=0, \nonumber\\
 v_x -T[u]=0, \nonumber\\
 G_\alpha [u]=0,\qquad \alpha =2,\ldots ,M. \label{eq27}
\end{gather}
Suppose a point symmetry (not in evolutionary form)
\[
X=\xi (x,t,u,v){\partial \over {\partial x}}+\tau (x,t,u,v){\partial \over {\partial t}}+\eta
^j(x,t,u,v){\partial \over {\partial u^j}}+\phi (x,t,u,v){\partial \over {\partial v}}
\]
is admitted by the equivalent potential system (\ref{eq27}). Then $X$ yields a \textit{nonlocal} symmetry of the given
system (\ref{eq25}) if and only if $(\xi _v )^2+(\tau _v )^2+\sum\limits_{j=1}^m {(\eta _v^j } )^2 \neq 0.$

Now suppose none of the equations of a given system of PDEs (\ref{eq25}) is written as a conservation law. Suppose the set
of multipliers $\{\Lambda ^\sigma [U]\}$ yields a conservation law of system (\ref{eq25}), i.e.,
\begin{gather}
\label{eq28} \Lambda ^\sigma [U]G_\sigma [U]=D_t T[U]+D_x X[U].
\end{gather}
Suppose for some $\beta, \Lambda ^\beta [U]\neq0$ and suppose every solution of the system $\Lambda ^\beta [u]=0$,
$G_\sigma [u]=0$, $\sigma =1,\ldots ,M$ with $\sigma \ne \beta ,$ is also a solution 
of the given system of PDEs (\ref{eq25}).
Then the conservation law (\ref{eq28}) is said to be a \textit{useful} 
conservation law of system (\ref{eq25}). In the case
of a~useful conservation law (without loss of generality $\beta =1),$ the given system of PDEs (\ref{eq25}) is non-locally
equivalent to a resulting potential system of the form (\ref{eq27}) since this equivalence follows when one replaces
$\Lambda ^1[u]G_1 [u]=0$ by the potential equations (27a,b).

\section{Connections between symmetries and conservation laws\\ for linearization}\label{sec6}

In \cite{Bluman&Kumei,Bluman&Kumei3}, the following two theorems are proven for determining whether or not a given nonlinear
system of PDEs (\ref{eq10}) can be invertibly mapped to some linear system of PDEs and to find such a mapping when it exists
in terms of admitted symmetries of the nonlinear system. Here we present the theorems when the number of dependent variables
$m\ge 2.$

\begin{theorem}[Necessary conditions for the existence of an invertible mapping]\label{theorem4} \textit{If there exists an invertible transformation that
maps a given nonlinear system of PDEs} \eqref{eq10} \textit{to some linear system of PDEs, then}
\begin{enumerate}
\itemsep=0pt
\item \textit{the mapping must be a point transformation of the form}
\begin{gather}
 z_j =\phi _j (x,u),\qquad j=1,2,\ldots ,n, \nonumber\\
 w^\gamma =\psi ^\gamma (x,u),\qquad \gamma =1,2,\ldots ,m;\label{eq29}
\end{gather}
\item \textit{the given system of PDEs} \eqref{eq10} \textit{must admit an infinite-parameter Lie group of point transformations having an infinitesimal generator }
\begin{gather}\label{eq30m}
X=\xi _i [u]{\partial \over {\partial x_i }}+\eta ^\nu [u]{\partial \over {\partial u^\upsilon }}
\end{gather}
\textit{with each} $\xi _i [u]$, $\eta ^\nu [u]$ \textit{characterized by}
\begin{gather}
\label{eq30} \xi _i [u]=\sum\limits_{\sigma =1}^m {\alpha _i^\sigma } (x,u)F^\sigma (x,u),\qquad \eta ^\nu
[u]=\sum\limits_{\sigma =1}^m {\beta _\nu ^\sigma } (x,u)F^\sigma (x,u),
\end{gather}
\textit{where} $\alpha _i^\sigma$, $\beta _\nu ^\sigma$ 
\textit{are specific functions of} $x$ and $u$, and $F=(F^1,F^2,\ldots
,F^m)$ \textit{is an arbitrary solution of some linear system of PDEs}
\begin{gather}
\label{eq31} \mbox{L}F=0
\end{gather}
\textit{with} $L$ \textit{representing a linear differential operator in terms of some independent variables}
$X=(X_1 (x,u),X_2 (x,u),\ldots ,X_n (x,u)).$
\end{enumerate}

\end{theorem}

\begin{theorem}[Sufficient conditions for the existence of an
invertible mapping]\label{theorem5}
{Suppose a given nonlinear system of PDEs} \eqref{eq10} \textit{admits an infinitesimal generator} \eqref{eq30m}
 \textit{whose
coefficients are of the form } \eqref{eq30} \textit{with F being an arbitrary solution of some linear system}
\eqref{eq31}
\textit{with specific independent variables }$X=(X_1 (x,u),X_2 (x,u),\ldots ,X_n (x,u)).$
\textit{If the linear homogeneous
system of m first order PDEs for the scalar function} $\Phi $\textit{ given by}
\begin{gather}
\label{eq32} \alpha _i^\sigma (x,u)\frac{\partial \Phi }{\partial x_i }+\beta _\nu ^\sigma (x,u)\frac{\partial \Phi
}{\partial u^\nu }=0,\qquad \sigma =1,2,\ldots ,m,
\end{gather}
\textit{has n functionally independent solutions}
$X_1 (x,u),X_2 (x,u),\ldots ,X_n (x,u)$, \textit{and the linear system of
}$m^2$\textit{ first order PDEs}
\begin{gather}
\label{eq33} \alpha _i^\sigma (x,u)\frac{\partial \psi ^\gamma }{\partial x_i }+\beta _\nu ^\sigma (x,u)\frac{\partial \psi
^\gamma }{\partial u^\nu }=\delta ^{\gamma \sigma },\qquad \gamma ,\sigma =1,2,\ldots ,m,
\end{gather}
\textit{where }$\delta ^{\gamma \sigma }$\textit{is the Kronecker symbol, has a solution }$\psi =(\psi ^1(x,u),\psi
^1(x,u),\ldots ,\psi ^m(x,u))$, \textit{then the invertible mapping given by}
\begin{gather*}
 z_j =\phi _j (x,u)=X_j (x,u),\qquad j=1,2,\ldots ,n, \\
 w^\gamma =\psi ^\gamma (x,u),\qquad \gamma =1,2,\ldots ,m, 
 \end{gather*}
\textit{transforms the nonlinear system of PDEs} \eqref{eq10} \textit{to the linear system of PDEs given by}
\[
\mbox{L}w=g(z)
\]
\textit{for some function} $g(z).$
\end{theorem}

The theorems for linearization through admitted infinite-parameter
groups have their counterparts for admitted multipliers
for conservation laws \cite{Bluman&DoranWu}. This follows first 
from the observation that for any linear operator $L$ and
its adjoint operator $\mbox{L}^\ast ,$ one has the well-known relationship
\begin{gather}
\label{eq34} V\mbox{L}U-U\mbox{L}^\ast V=D_i P^i[U,V]
\end{gather}
for some $P^i[U,V]$ that depend bilinearly 
on any functions $U(x),\;V(x)$ and their derivatives to some finite order. Hence
if a given system of PDEs (\ref{eq10}) is linear, i.e., it is of the form
\begin{gather}
\label{eq35} \mbox{L}u=0,
\end{gather}
then \textit{any} set of multipliers $\{\Lambda ^\sigma (x)\}$ satisfying
\begin{gather}
\label{eq36} \mbox{L}^\ast \Lambda (x)=0
\end{gather}
yields a conservation law for system (\ref{eq34}).

The above observation combined with the result that a point transformation 
(\ref{eq29}) maps any conservation law of a given
system of PDEs into a conservation law of another system of PDEs in terms of the variables of the point transformation
\cite{Bluman&Temuerchaolu&Anco}, shows that if a given nonlinear system of PDEs (\ref{eq10}) 
is linearizable by a point
transformation (\ref{eq29}), then it is necessary that (\ref{eq10}) admits a set of multipliers
 {\{}$\Lambda ^\sigma [U]\}$
of the form
\begin{gather}
\label{eq37} \Lambda ^\sigma [U]=A_\rho ^\sigma [U]F^\rho (X);
\end{gather}
$A_\rho ^\sigma$ are specific functions of $x$, $U$ and derivatives of $U$ to some finite order, and 
$X=(X_1 (x,U)$, $X_2
(x,U),\ldots ,X_n (x,U))$ are specific functions of $x$ and $U$ with the property that if $U=u$ is any solution of the given
nonlinear system (\ref{eq10}), then for some linear operator $\mbox{L}^\ast $ in terms of independent variables $z=X=(X_1
(x,u),X_2 (x,u),\ldots ,X_n (x,u)),$ one has
\begin{gather}
\label{eq38} \mbox{L}^\ast F=0.
\end{gather}
If a given nonlinear system (\ref{eq10}) can be mapped by a point transformation (\ref{eq29}) into a linear system
$\mbox{L}w=0$, then the operator $\mbox{L}^\ast $ in (\ref{eq38}) is the adjoint of L.

The results presented in this section extend to contact transformations in the case of a scalar DE
\cite{Bluman&Kumei,Bluman&DoranWu,Kumei&Bluman2,Bluman&Kumei3,Bluman&Kumei4,Anco&Wolf&Bluman}.

In summary, if a given nonlinear system of PDEs can be mapped invertibly into a linear system, then its admitted symmetries
yield the linear system and the mapping to the linear system whereas its admitted multipliers yield the adjoint linear
system. Further details on linearization through conservation laws will appear in \cite{Anco&Wolf&Bluman}.

\section{Examples}\label{sec7}

Now we illustrate all of the above connections between symmetries and conservation laws through several examples involving
the nonlinear telegraph (NLT) equation
\begin{gather}
\label{eq39} u_{tt} -(F(u)u_x )_x -(G(u))_x =0.
\end{gather}
For \textit{any} $(F(u),G(u))$ pair, from PDE (\ref{eq39}) written as a conservation law, we obtain potential systems
\begin{gather}
 G_1 [u,v]=v_t -F(u)u_x -G(u)=0,\nonumber \\
 G_2 [u,v]=v_x -u_t =0; \label{eq40}\\
 H_1 [u,v,w]=G_1 [u,v]=0, \nonumber\\
 H_2 [u,v,w]=w_t -v=0, \nonumber\\
 H_3 [u,v,w]=w_x -u=0.\label{eq41}
\end{gather}
The point symmetry classification of the NLT scalar equation (\ref{eq39}) appears in \cite{Anco&Wolf&Bluman}. 
The point symmetry classification
of the potential system (\ref{eq40}) yields nonlocal symmetries for the NLT equation (\ref{eq39}) for a very wide class of
$(F(u),G(u))$ pairs \cite{Bluman&Temuerchaolu&Sahadevan}. For \textit{specific }$(F(u),G(u))$ pairs, the conservation law
classification results for potential system (\ref{eq40}) yield additional useful conservation laws and an extended tree of
non-locally related systems \cite{Bluman&Cheviakov, Bluman&Temuerchaolu1}. For the potential system~(\ref{eq40}), it is
interesting to compare its conservation law and point symmetry classifications~\cite{Bluman&Temuerchaolu2}.

\subsection{Nonlocal symmetries}\label{ss71}

The point symmetry
\begin{gather}
\label{eq42} X=\xi (x,t,u,v){\partial \over {\partial x}}+\tau (x,t,u,v){\partial \over {\partial
t}}+\eta (x,t,u,v){\partial \over {\partial u}}+\phi (x,t,u,v){\partial \over {\partial v}}
\end{gather}
is admitted by the potential system (\ref{eq40}) if and only if the system of determining equations
\begin{gather*}
 \xi _v -\tau _u =0, \\
 \eta _u -\phi _v +\xi _x -\tau _t =0, \\
 G(u)[\eta _v +\tau _x ]+\eta _t -\phi _x =0, \\
 \xi _u -F(u)\tau _v =0, \\
 \phi _u -G(u)\tau _u -F(u)\eta _v =0, \\
 G(u)\xi _v +\xi _t -F(u)\tau _x =0, \\
 F(u)[\phi _v -\tau _t +\xi _x -\eta _u -2G(u)\tau _v ]-{F}'(u)\eta =0, \\
 G(u)[\phi _v -\tau _t -G(u)\tau _v ]-F(u)\eta _x -{G}'(u)\eta +\phi _t =0,
 \end{gather*}
holds for \textit{arbitrary }values of $x$,
$t$, $u$, $v.$ One can show \cite{Bluman&Temuerchaolu&Sahadevan} that the potential
system (\ref{eq40}) yields a nonlocal symmetry of the NLT equation (\ref{eq39}) if and only if
\begin{gather}
 (c_3 u+c_4 ){F}'(u)-2(c_1 -c_2 -G(u))F(u)=0, \nonumber\\
 (c_3 u+c_4 ){G}'(u)+G^2(u)-(c_1 -2c_2 +c_3 )G(u)-c_5 =0,\label{eq43} 
\end{gather}
for arbitrary constants $c_i$, $i=1,\ldots ,5.$ 
Moreover, the potential system (\ref{eq40}) allows (\ref{eq39}) to be
linearizable (by a nonlocal transformation) if and only if in (\ref{eq43}), one has $c_1 =0$,
$c_5 =c_2 (c_3 -c_2 ).$ For
any $(F(u),G(u))$ pair satisfying ODE system (\ref{eq43}), 
the potential system (\ref{eq40}) admits a point symmetry of the
form (\ref{eq42}) with
\begin{gather*}
 \xi =c_1 x+\int {F(u)du,} \\
 \tau =c_2 t+v, \\
 \eta =c_3 u+c_4 , \\
 \phi =c_5 t+(c_1 -c_2 +c_3 )v, 
 \end{gather*}
which in turn yields a nonlocal symmetry admitted by the given NLT equation (\ref{eq39}) since $\tau $ has an essential
dependence on $v$.

\subsection{Conservation laws}\label{ss72}

The multipliers $[\Lambda _1 ,\Lambda _2 ]=[\alpha (x,t,U,V),\beta (x,t,U,V)]$ yield a conservation law of the NLT potential
system (\ref{eq40}) if and only if for the Euler operators $E_U$, $E_V$ one has
\begin{gather}
 E_U (\alpha G_1 [U,V]+\beta G_2 [U,V])\equiv 0, \nonumber\\
 E_V (\alpha G_1 [U,V]+\beta G_2 [U,V])\equiv 0, \label{eq44}
\end{gather}
for \textit{arbitrary }differentiable functions $U(x,t)$,
$V(x,t).$ Equations (\ref{eq44}) reduce to the following determining
equations for the multipliers:
\begin{gather}
 \beta _V -\alpha _U =0, \nonumber\\
 \beta _U -F(U)\alpha _V =0, \nonumber\\
 \beta _x -\alpha _t -G(U)\alpha _V =0, \nonumber\\
 F(U)\alpha _x -\beta _t -[G(U)\alpha ]_U =0. \label{eq45}
\end{gather}
For any solution of the determining equations (\ref{eq45}), 
one can show that the corresponding conserved densities are
given by
\begin{gather*}
X=- \int_{a}^{u} {\alpha (x,t,s,b)ds} -\int_{b}^{v} {\beta
(x,t,u,s)ds}-G(a)\int^{x} {\alpha (s,t,a,b)ds}, \\
T=\int_{a}^{u} \beta (x,t,s,b)ds+\int_{b}^{v} \alpha
(x,t,u,s)ds, 
\end{gather*}
in terms of arbitrary constants $a$ and $b$.

In solving the determining equations (\ref{eq45}), it turns out that three cases arise in terms of classifying functions
$d(U)$ and $h(U)$ given by
\begin{gather*}
 d(U)={G}'^2{F}'''-3{G}'{G}''{F}''+(3{G}''^2-{G}'{G}'''){F}', \\
 h(U)={G}'^2G^{(\ref{eq4})}-4{G}'{G}''{G}'''+3{G}''^3. 
 \end{gather*}
In particular, one must separately consider the cases 
$d(U)=h(U)= 0;$ $d(U)\neq0,~h(U)\equiv 0$; $d(U)\neq0,~h(U)\neq0.$ For
further details, see \cite{Bluman&Temuerchaolu1}.

\begin{example}
As a first example, consider the NLT potential system
\begin{gather}
 v_t +(1-2e^{2u})u_x -e^u=0,\nonumber \\
 v_x -u_t =0. \label{eq46}
\end{gather}
After solving the corresponding determining equations (\ref{eq43}), 
one can show that potential system~(\ref{eq46}) admits
the set of multipliers
\begin{gather}
 \alpha =e^{-{1 \over 2}(U+t/\sqrt 2 )}\sin \left({1 \over
2}\left(V+\frac{x+2e^U}{\sqrt 2}\right)\right), \nonumber\\
 \beta =-e^{-{1 \over 2}(U+t/\sqrt 2 )}\left[\sqrt 2 e^U\sin
\left({1 \over 2}\left(V+\frac{x+2e^U}{\sqrt 2}\right)\right)+\cos \left({1 \over
2}\left(V+\frac{x+2e^U}{\sqrt 2}\right)\right)\right]\label{eq47}
\end{gather}
with corresponding densities
\begin{gather}
 T=-2e^{-{1 \over 2}(u+t/\sqrt 2 )}\cos \left({1 \over
2}\left(v+\frac{x+2e^u}{\sqrt 2}\right)\right), \nonumber\\
 X=2e^{-{1 \over 2}(u+t/\sqrt 2 )}\left[\sqrt 2 e^u\cos \left({1
\over 2}\left(v+\frac{x+2e^u}{\sqrt 2}\right)\right)-\sin \left({1 \over 2}\left(v+\frac{x+2e^u}{\sqrt
2}\right)\right)\right]. \label{eq48}
\end{gather}

\end{example}

\begin{example}As a second example, consider the NLT potential system
\begin{gather}
 v_t -(\mbox{sech}^2u)u_x +\tanh u=0, \nonumber\\
 v_x -u_t =0. \label{eq49}
\end{gather}
The potential system (\ref{eq49}) admits the set of multipliers
\begin{gather}
\label{eq50} \alpha =e^x(2x+t^2-V^2-2\log (\cosh U)),\qquad \beta =2e^x(V\tanh U-t)
\end{gather}
with corresponding densities
\begin{gather}
 T=e^x\left[2tu-{1 \over 3}v^3+v(t^2+2x-2\log (\cosh u))\right],\nonumber \\
 X=e^x\left[(v^2-t^2-2x+2(1+\log (\cosh u)))\tanh u-2(tv+u)\right].\label{eq51}
\end{gather}
\end{example}

\subsection{Symmetry action on conservation laws to yield new conservation laws}\label{ss73}

\begin{example} The NLT potential system (\ref{eq46}) obviously admits the discrete reflection symmetry
\begin{gather}
\label{eq52} (t,x,u,v)=(-\tilde {t},\tilde {x},\tilde {u},-\tilde {v})
\end{gather}
and the one-parameter $(\varepsilon)$ family of translations
\begin{gather}
\label{eq53} (t,x,u,v)=(\tilde {t},\tilde {x},\tilde {u},\tilde {v}+\varepsilon ).
\end{gather}
Applying these symmetries to (\ref{eq47}), (\ref{eq48}),
 one is able to find three more conservation laws for system~(\ref{eq46}).

The reflection symmetry (\ref{eq52}) applied to (\ref{eq47}) yields a second set of multipliers
\begin{gather}
\label{eq54} [\alpha _2 ,\beta _2 ]=[-\alpha (x,-t,U,-V),\beta (x,-t,U,-V)]
\end{gather}
with corresponding densities
\begin{gather}
\label{eq55} [T_2 ,X_2 ]=[T(x,-t,u,-v),-X(x,-t,u,-v)].
\end{gather}
The translation symmetry (\ref{eq53}) applied to (\ref{eq47}), (\ref{eq48}) yields a third set of multipliers (from the
$O(\varepsilon )$ term in the corresponding expression (\ref{eq24}))
\begin{gather}
\label{eq56} [\alpha _3 ,\beta _3 ]=[\alpha (x,t,U,V+\pi ),\beta (x,t,U,V+\pi )]
\end{gather}
and conserved densities
\begin{gather}
\label{eq57} [T_3 ,X_3 ]=[T(x,t,u,v+\pi ),X(x,t,u,v+\pi )].
\end{gather}
Finally, the reflection symmetry (\ref{eq52}) applied to (\ref{eq56}), (\ref{eq57}) yields a fourth set of multipliers
$[\alpha _4 ,\beta _4 ]=[-\alpha (x,-t,U,-V-\pi ),\beta (x,-t,U,-V-\pi )]$ with corresponding densities $[T_4 ,X_4
]=[T(x,-t,u,-v-\pi ),-X(x,-t,u,-v-\pi )].$
\end{example}


\begin{example} The NLT potential system (\ref{eq49}) obviously admits the one-parameter $(\varepsilon)$ family of translations
\begin{gather}
\label{eq58} (t,x,u,v)=(\tilde {t}+\varepsilon ,\tilde {x},\tilde {u},\tilde {v})
\end{gather}
as well as the point symmetry associated with the infinitesimal generator
\begin{gather}
\label{eq59} X=v{\partial \over {\partial t}}+\tanh u{\partial \over {\partial x}}+{\partial
\over {\partial u}}+t{\partial \over {\partial v}}.
\end{gather}
Applying these two symmetries to (\ref{eq50}), (\ref{eq51}), one is able to generate three more conservation laws as
follows.

From the $O(\varepsilon )$ and $O(\varepsilon ^2)$ terms that result from the corresponding expression (\ref{eq24}) 
when~(\ref{eq58}) is applied to (\ref{eq50}), (\ref{eq51}), one respectively obtains two new sets of 
multipliers $[\alpha _2
,\beta _2 ]=[te^x,-e^x]$, $[\alpha _3 ,\beta _3 ]=[e^x,0]$ and conserved densities $[T_2 ,X_2 ]=[e^x(tv+u),-e^x(v+t\tanh
u)]$, $[T_3 ,X_3 ]=[e^xv,-e^x\tanh u].$ The $O(\varepsilon )$ term that arises 
from expression (\ref{eq24}) when symmetry
(\ref{eq59}) is applied to $[\alpha _2 ,\beta _2 ]$, $[T_2 ,X_2 ]$ yields a fourth set 
of multipliers $[\alpha _4 ,\beta _4
]=[e^xV,-e^x\tanh U]$ 
and correspon\-ding conserved densities $[T_4 ,X_4 ]=\big[e^x\big({1 \over 2}v^2+\log (\cosh
u)\big),-e^xv\tanh u\big].$
\end{example}

\subsection{Linearization}\label{ss74}

Now we consider two examples of NLT potential systems that can be linearized and see how this can be done from admitted
symmetries and observed from admitted multipliers for conservation laws.

\begin{example}The quasilinear NLT potential system
\begin{gather}
 v_t -F(u)u_x =0, \nonumber\\
 v_x -u_t =0, \label{eq60}
\end{gather}
admits the point symmetry $X=A(u,v){\partial \over {\partial x}}+B(u,v){\partial \over {\partial t}}$
with
\begin{gather}
 A_u -F(u)B_v =0,\nonumber \\
 A_v -B_u =0.\label{eq61}
\end{gather}
In the symmetry Theorems 1 and 2, let $x_1 =t$,
$x_2 =x$, $u^1=u$, $u^2=v.$ Then the conditions in Theorem 1 yield
$F^1=B(u,v)$, $F^2=A(u,v)$,
$\alpha _1^1 =\alpha _2^2 =1$,
$\alpha _2^1 =\alpha _1^2 =0$,
$\beta _\nu ^\sigma =0$
with the linear system given by (\ref{eq61}) for $(X_1 ,X_2 )=(u,v).$ 
Here system (\ref{eq32}) becomes $\Phi _t =\Phi _x =0$
with functionally independent solutions $X_1 =u$,
$X_2 =v;$ system (\ref{eq33}) becomes ${{\partial \psi ^1}
\over {\partial t}}={{\partial \psi ^2} \over {\partial x}}=1,$ ${{\partial \psi ^1} \over {\partial
x}}={{\partial \psi ^2} \over {\partial t}}=0$ and has as a solution $\psi ^1=t$,
$\psi ^2=x.$ This yields the
well-known hodograph transformation $z_1 =u$, $z_2 =v$,
$w^1=t$, $w^2=x$ that maps (\ref{eq60}) to the linear system
$x_u -F(u)t_v =0$, $x_v -t_u =0.$

On the other hand, system (\ref{eq60}) admits 
multipliers $[\Lambda ^1,\Lambda ^2]=[a(U,V),b(U,V)]$ satisfying the linear
system of PDEs
\begin{gather}
 b_V +a_U =0, \nonumber\\
 b_U +F(U)a_V =0. \label{eq62}
\end{gather}
Observe that system (\ref{eq62}) is the adjoint of system (\ref{eq61}).
\end{example}

\begin{example}The NLT potential system
\begin{gather}
 v_t -u^{-2}u_x -u^{-1}=0, \nonumber\\
 v_x -u_t =0, \label{eq63}
\end{gather}
admits the point symmetry $X=-u^{-1}A(\hat {u},v){\partial \over {\partial x}}+B(\hat {u},v){\partial
\over {\partial t}}+A(\hat {u},v){\partial \over {\partial u}}$, $\hat {u}=x+\log u,$ with
\begin{gather}
 A_v +B_{\hat {u}} =0, \nonumber\\
 A_{\hat {u}} +B_v -A=0. \label{eq64}
\end{gather}
Here the conditions in Theorem 1 yield $F^1=B(\hat {u},v)$,
$F^2=A(\hat {u},v)$,
$\alpha _1^1 =1$,
$\alpha _2^2 =-u^{-1}$,
$\alpha _1^2 =\alpha _2^1 =0$,
$\beta _1^2 =1$,
$\beta _1^1 =\beta _2^2 =\beta _2^1 =0$ with the linear system
given by (\ref{eq64}) for $(X_1 ,X_2 )=(\hat {u}=x+\log u,v).$ System (\ref{eq32}) becomes $\Phi _t =0$,
$\Phi _u -u^{-1}\Phi _x =0$ with functionally independent solutions $X_1 =\hat {u}$,
$X_2 =v;$ system (\ref{eq33}) becomes
${{\partial \psi ^1} \over {\partial t}}=1$,
${{\partial \psi ^1} \over {\partial
u}}-u^{-1}{{\partial \psi ^1} \over {\partial x}}=0,{{\partial \psi ^2} \over {\partial
t}}=0$, ${{\partial \psi ^2} \over {\partial u}}-u^{-1}{{\partial \psi ^2} \over {\partial x}}=1$ and
has as a solution $\psi ^1=t$, $\psi ^2=u.$ This yields the transformation $z_1 =\hat {u}=x+\log u$,
$z_2 =v$,
$w^1=t$,
$w^2=u$ that maps (\ref{eq63}) to the linear system ${{\partial w^2} \over {\partial z_2
}}+{{\partial w^1} \over {\partial z_1 }}=0,\;\;{{\partial w^2} \over {\partial z_1
}}+{{\partial w^1} \over {\partial z_2 }}-w^2=0.$

On the other hand, system (\ref{eq63}) admits multipliers for conservation 
laws given by $[\Lambda ^1,\Lambda ^2]=[Ua(\hat
{U},V),b(\hat {U},V)], \quad \hat {U}=x+\log U,$ with
\begin{gather}
 a_V +b_{\hat {U}} =0, \nonumber\\
 a_{\hat {U}} +b_V +a=0.\label{eq65}
\end{gather}
Again observe that system (\ref{eq65}) is the adjoint of system (\ref{eq64}).
\end{example}

\subsection*{Acknowledgements}

The author thanks his collaborators for much of the work presented 
in this paper, especially Stephen Anco, Alexei Cheviakov,
Sukeyuki Kumei and Temuerchaolu. He also acknowledges 
financial support from the National Sciences and Engineering Research
Council of Canada.

\LastPageEnding

\end{document}